\documentclass[11pt]{article}
\usepackage{amsmath}
\usepackage{graphicx,epsfig}

\begin{document}





\title{Action in the Entropic Revolution of Newtonian Gravity}

\author{Joakim Munkhammar \footnote{Studentstaden 23:230, 752 33
Uppsala, Sweden, joakim.munkhammar@gmail.com}}
\date{\today}
\maketitle

\begin{abstract}
The theory of gravity has undergone somewhat of a revolution
lately. Gravity is no longer a fundamental force it seems, but
rather an effect of holographic entropy. Building on the works by
Jacobsson, Padmanabhan and Verlinde we review the concept of
Newtonian gravity as an entropic force and discuss a possible
general action approach to Verlinde's theory. We also discuss some
open problems and future prospects of Verlinde's
approach.


\end{abstract}

\vspace{10pt}

\begin{center}
\begin{small}
\textit{Essay written for the Gravity Research Foundation 2010
Awards for Essays on Gravitation}
\end{small}
\end{center}

\section{Introduction}
In a remarkable paper Erik Verlinde recently proposed a framework
for gravity as an entropic force \cite{Verlinde}. This theory,
which was built on the works of Jacobsson \cite{Jacobsson} and
Padmanabhan \cite{Padmanabhan,Padmanabhan2,Padmanabhan3}, showed
that Newtonian gravity could be obtained with very simple tools
and utilizing practically only one assumption. His assumption was
that the holographic principle holds and that space is emergent
\cite{Verlinde}. Perhaps the most important concept he used was
the concept of holographic entropy, which was discovered in the
1970s by Bekenstein \cite{Bekenstein}, as the defining component
of the gravitational force \cite{Verlinde}. Thoughts similar to
Verlinde's had been put forward before by Jacobsson
\cite{Jacobsson} and Padmanabhan \cite{Padmanabhan}, where they
utilized Rindler space time and reverse-engineered the framework
to get the Einstein field equations as well as the
Lanchos-Lovelock equations for gravity \cite{Padmanabhan}. This
research entirely reversed the view on physics and in particular
the view on gravity. All of a sudden the fundamental laws were the
result of a deeper theory that had been derived from a more
advanced one, this almost rendered physics either circular or a
\textit{strange loop} \cite{Hofstadter}. Verlinde concluded that
the change of entropy was linked to the change of the Newtonian
potential, this led to the conclusion that inertia might be
equivalent to the lack of entropy gradients \cite{Verlinde}. He
states the following in his seminal paper \cite{Verlinde}:

\vspace{10pt}

\textit{The holographic principle has not been easy to extract
from the laws of Newton and Einstein, and is deeply hidden within
them. Conversely, starting from holography, we find that these
well known laws come out directly and unavoidably.}

\vspace{10pt}

His paper attracted quite some attention and several papers from
various fields of theoretical physics, including cosmology, Loop
Quantum Gravity and quantum mechanics, have been published
relating to its topic
\cite{Easson,Ghosh,Munkhammar,Smolin,Vancea,Zhao}. We shall
breifly review some parts of Verlinde's theory and discuss a
possible action for his theory.




\section{The holographic principle}
In the holographic view space is mainly a storage place of
information, which is associated with positions, movements and
mass of matter
\cite{Bousso,Munkhammar,Padmanabhan,Padmanabhan2,Susskind,tHooft,Verlinde}.
This information is displayed on a surface, a holographic screen
\cite{Verlinde}. The information is stored in discrete bits on the
screen and since the number of bits is limited we get holographic
effects. This is the holographic principle in practice. Thus the
dynamics on the screen is governed by some unknown rules which
then only can utilize the information on the screen. Since
information is stored on a screen this means that space is
emergent in the normal direction of the screen \cite{Verlinde}.
The microstates may be thought of having all sorts of physical
attributes such as energy, temperature etc. This is then related,
via entropy, to the information associated with the system
\cite{Verlinde}.

\section{Entropy as a gravitational force}
Bekenstein related the area of a black hole to the entropy of it
by assuming that all information lost down a black hole must still
be conserved and is therefore contained in some measure
\cite{Bekenstein,Hawking}. If we now in the gravitational
situation consider a small piece of a holographic screen and a
particle with mass $m$ approaching it from the side at which time
has already emerged, then Verlinde concluded (unitilzing
Bekenstein's arguments) that the entropy associated with this
process should be Bekenstein entropy with an extra factor of
$2\pi$ \cite{Verlinde}:
\begin{equation}\label{EntropicForce}
\Delta S = 2 \pi k_B \frac{mc}{\hbar} \Delta x.
\end{equation}
Here $k_B$ is Boltzmanns constant and the factor $2\pi$ was added
by Verlinde for reasons to be clear in connection with the
gravitational force. An entropic force $F$ is defined as
\cite{Verlinde}:
\begin{equation}\label{Force}
F \Delta x = T \Delta S
\end{equation}
where $T$ is temperature. It should also be noted that an entropic
force is a macroscopic force that originates in a system with many
degrees of freedom by the universe's statistical tendency to
maximize its entropy \cite{Verlinde}. In order to relate the
entropy to the screen the maximum number of bits $N$ that can be
associated with a screen is then assumed to be:
\begin{equation}\label{NumberofBits}
N = \frac{A c^3}{G \hbar} = \frac{4 \pi R^2 c^3}{G \hbar},
\end{equation}
where $R$ is the radius and $A = 4\pi R^2$ is the area of the
screen. The temperature can be determined from the equipartition
rule:
\begin{equation}\label{BitEnergy}
E = \frac12 N k_B T,
\end{equation}
which is the the average energy per bit on the screen
\cite{Padmanabhan,Padmanabhan2}. We shall also assume that the
mass-energy relation holds:
\begin{equation}\label{EMC}
E = Mc^2.
\end{equation}
In a straight forward way these equations yields the gravitational
force:
\begin{equation}
F = G \frac{Mm}{R^2}.
\end{equation}
This is, as Verlinde points out, a surprising result considering
it practically comes from first principles \cite{Verlinde}.
Verlinde also showed another way to set up the force of gravity as
an entropic force by identifying the Unruh temperature arising in
Rindler space time as the result of the gradient of the Newtonian
potential $\phi$ \cite{Verlinde}:
\begin{equation}\label{Unruh}
k_B T = \frac{1}{2\pi} \frac{\hbar a}{c} = \frac{1}{2\pi}
\frac{\hbar \nabla \phi}{c}.
\end{equation}
The number of bits on a screen \eqref{NumberofBits} can be put in
a differential form:
\begin{equation}\label{Bitsperarea}
dN = \frac{c^3}{G \hbar} dA.
\end{equation}
The equipartition rule \eqref{BitEnergy} can also be formulated on
integral form:
\begin{equation}\label{Equipartition2}
E = \frac12 k_B \int_\mathcal{S} T dN,
\end{equation}
where $\mathcal{S}$ is the screen enclosing the particle. If we
just insert \eqref{Bitsperarea} and \eqref{EMC} in
\eqref{Equipartition2} and use the Unruh temperature \eqref{Unruh}
we get the expression:
\begin{equation}\label{Gauss}
M = \frac{1}{4 \pi G} \int_\mathcal{S} \nabla \phi \cdot dA.
\end{equation}
This is Gauss law of gravity. Verlinde points out that this should
hold for arbitrary screens given by equipotential surfaces
$\mathcal{S}$ \cite{Verlinde}. Gauss law of gravity also
transforms into Poisson's equation for gravity:
\begin{equation}
\nabla^2 \phi = 4 \pi G \rho.
\end{equation}
Verlinde then utilized Killing vectors in order to obtain a
relativistic version of the field, an approach which was similar
to the previous approach by Jacobsson \cite{Jacobsson}. For strong
fields the relativistic version of the theory turns out to be
equivalent to Einstein field equations, see
\cite{Jacobsson,Padmanabhan,Verlinde} for more information.
Verlinde's general conclusion is that inertia is due to the lack
of entropy gradients, and conversely that gravity is due to the
presence of them \cite{Verlinde}.


\section{Action formulation of Verlinde's gravity}
In the context of Verlinde's theory any physical theory will
posses gravity naturally and unavoidably if there are entropy
gradients present \cite{Verlinde}. This means in effect that the
action should be based on a Lagrangian that has the extra term of
entropic energy. We have the energy from the entropic force on the
integral form:
\begin{equation}
E = \int F dx = \int T dS.
\end{equation}
The entropic energy should be considered a potential energy. Thus
any physical system described by the Lagrangian $\mathcal{L}$ that
does not \textit{a priori} contain gravity (and it should not) is
then reexpressed with gravity by the Lagrangian $\mathcal{L}'$:
%
%
\begin{equation}
\mathcal{L}' = \mathcal{L} + \int T dS.
\end{equation}
This amounts to the action $\mathcal{A}$ when integrated over
time:
\begin{equation}
\mathcal{A} = \int \Bigg(\mathcal{L} + \int TdS \Bigg)dt.
\end{equation}
This approach to action from the entropic force is akin to
Padmanabhan's approach \cite{Padmanabhan3}, but here applied
directly to Verlinde's non-relativistic theory \cite{Verlinde}.
The extra addition of the entropic force energy to the action may
be seen as surface terms in many field theories, which are ignored
in the conventional approach \cite{Padmanabhan3}. If we assume the
differential form of the entropy equation \eqref{EntropicForce} by
letting $\Delta S \to dS$ then we can for a small infalling mass
$m$ with the use of the temperature from the equipartition law
\eqref{BitEnergy} compute the entropy energy:
\begin{equation}
\int TdS = -\frac{GMm}{r} = - m\phi.
\end{equation}
This is equal to the Newtonian potential energy with negative
sign. If one works out the variational principle ($\delta
\mathcal{A} =0$) \cite{Marion} for the single particle with
$\mathcal{L} = \frac12 m v^2$ one naturally obtains Newton's law
of gravity in this approach:
\begin{equation}
m \ddot{x} = - m\nabla \phi.
\end{equation}
The role of the action in the relativistic version of Verlinde's
theory is left open. Such a generally covariant action should
correspond to the Einstein-Hilbert action of general relativity
for strong fields \cite{Jacobsson,Padmanabhan3,Verlinde}. The weak
field limit of such an action, just as in Verlinde's field
equations, should then be of particular interest \cite{Verlinde}.


\section{Discussion}
There are many open problems and areas in need of investigation in
Verlinde's view of gravity. One may conclude that many papers have
already been published on the basis of Verlinde's theory, and
among these papers a number of approaches to a quantum mechanical
origin of holographic entropy have been proposed
\cite{Ghosh,Lee,Munkhammar,Vancea}. Indeed, it is perhaps the
quantum dominated situations in physics that will provide the most
interesting results in Verlinde's gravitational theory. The search
for a viable such theory is still ongoing, but the direction of
research is different than in most quantum gravity theories, in
fact Jacobsson states the following \cite{Jacobsson}:

\vspace{10pt}

\textit{This perspective suggests that it may be no more
appropriate to canonically quantize the Einstein equation than it
would be to quantize the wave equation for sound in air.}

\vspace{10pt}

Thus the quantum theory of gravity is perhaps just quantum
mechanics with the addition of gravity as a potential caused by
the entropic force \cite{Ghosh,Lee,Munkhammar,Vancea}. Another
interesting feature of Verlinde's theory ought to be the general
nature of the weak gravitational field, since his relativistic
field equations only correspond to the Einstein field equations
for strong fields. Perhaps the investigation of gravitomagnetism
as a form of post-newtonian approximation could provide
interesting observables. In addition to this one might finally
conclude the true nature of gravitational radiation and how it may
accurately be observed with such a linearized theory.



\begin{thebibliography}{MMMM}
\bibitem{Bekenstein}
J.D.Bekenstein, \textit{Black holes and entropy}, Phys. Rev. D
\textbf{7}, 2333 1973.

\bibitem{Bousso}
R.Bousso, \textit{The holographic principle},
arXiv:hep-th/0203101v2 2002.


\bibitem{Caravelli}
F.Caravelli, L.Modesto, \textit{Holographic actions from black
hole entropy}, arXiv:1001.4364v2 2010.

\bibitem{Easson}
D.A.Easson, P.H.Frampton, G.F.Smoot, \textit{Entropic Accelerating
Universe}, arXiv:1002.4278v1 [ hep-th] 2010.




\bibitem{Ghosh}
S.Ghosh, \textit{Planck Scale Effect in the Entropic Force Law},
arXiv:1003.0285 [hep-th] 2010.

\bibitem{Hawking}
S.W.Hawking, \textit{Particle creation by black holes}, Commun.
Math. Phys. 43, 199-220, 1975.

\bibitem{Hofstadter}
D.Hofstadter, \textit{I am a strange loop}, Basic books 2007.

\bibitem{Jacobsson}
T.Jacobsson, \textit{Thermodynamics of Spacetime: The Einstein
Equation of State}, arXiv:gr-qc/9504004v2 1995.

\bibitem{Konoplya}
R.A.Konoplya, \textit{Entropic force, holographyc and
thermodynamics for static space-times}, arXiv:1002.2818v3 [hep-th]
2010.

\bibitem{Lee}
J-W.Lee, H-C.Kim, J.Lee, \textit{Gravity as Quantum Entanglement
Force}, arXiv:1002.4568v1 [hep-th] 2010.


\bibitem{Marion}
J.B.Marion, S.T.Thornton, \textit{Classical Dynamics of Particles
and Systems}, fourth ed. Harcourt inc. 1995.

\bibitem{Munkhammar}
J.D.Munkhammar, \textit{Is Holographic Entropy and Gravity the
result of Quantum Mechanics?}, arXiv:1003.1262v2 [hep-th] 2010.




\bibitem{Padmanabhan}
T.Padmanabhan, \textit{Equipartition of energy in the horizon
degrees of freedom and the emergence of gravity}, arXiv:0912.3165
2009.

\bibitem{Padmanabhan2}
T.Padmanabhan, \textit{Gravitational Entropy Of Static Space-Times
And Microscopic Density Of States}, Class. Quant. Grav.
21:4485-4494, arXiv:gr-qc/0308070 2004.

\bibitem{Padmanabhan3}
T.Padmanabhan, \textit{Thermodynamical Aspects of Gravity: New
insights}, arXiv:0911.5004v2 [gr-qc] 2009.

\bibitem{Smolin}
L.Smolin, \textit{Newtonian gravity in loop quantum gravity},
arXiv:1001.3668v1 [gr-qc] 2010.

\bibitem{Susskind}
L.Susskind, \textit{The World as a Hologram},
arXiv:hep-th/9409089v2 1994.

\bibitem{tHooft}
G.'t Hooft, \textit{Dimensional reduction in quantum gravity},
arXiv:gr-qc/9310026v2 2009.

\bibitem{Vancea}
I.V.Vancea, M.A.Santos, \textit{Entropic Force Law, Emergent
Gravity and the Uncertainty Principle}, arXiv:1002.2454 [hep-th]
2010.

\bibitem{Verlinde}
E.Verlinde, \textit{On the Origin of Gravity and the Laws of
Newton}, arXiv:1001.0785v1 [hep-th] 2010.

\bibitem{Zhao}
Y.Zhao, \textit{Entropic force and its fluctuation from
gauge/gravity duality}, arXiv:1002.4039 [hep-th] 2010.


\end{thebibliography}
\end{document}